\documentclass[10pt]{article} 

\usepackage[top=2.54cm, bottom=2.54cm, left=2.75cm, right=2.75cm]{geometry} 

\usepackage{graphicx} 
\usepackage{mathtools}

\usepackage{cite}         
\newcommand{\citet}{\cite} 
\usepackage{color} 

\usepackage[T1]{fontenc}
\usepackage{graphicx}
\usepackage{mathtools}

\usepackage[utf8]{inputenc}
\usepackage{amssymb,amsmath}

\usepackage{amsfonts} 

\newcommand{\Z}{\mathbb{Z}}

\usepackage[nodisplayskipstretch]{setspace}
\usepackage{pdfpages}

\usepackage[pdfusetitle]{hyperref}
\usepackage[all]{xy}
\usepackage{graphicx}
\newcommand{\nc}{\newcommand}
\nc{\mbb}{\mathbb}\nc{\bb}{\mathbb}
\nc{\mbf}{\mathbf}\nc{\mb}{\mathbf}
\nc{\mc}{\mathcal}
\nc{\msf}{\mathsf}\nc{\ms}{\mathsf}
\nc{\acc}{\ms{acc}}
\nc{\ack}{\ms{ack}}
\nc{\alp}{\alpha}\nc{\al}{\alpha}\nc{\gka}{\alpha}
\nc{\ap}{\ms{ap}}
\nc{\apd}{\ms{apd}}
\nc{\base}{\ms{base}}\nc{\ba}{\ms{base}}
\nc{\bet}{\beta}\nc{\gkb}{\beta}
\nc{\boucle}{\ms{loop}}\nc{\Loop}{\ms{loop}}\nc{\lo}{\ms{loop}}
\nc{\bu}{\bullet}
\nc*{\cc}{\raisebox{-3pt}{\scalebox{2}{$\cdot$}}}
\nc{\centre}{\ms{center}}\nc{\Center}{\ms{center}}\nc{\cen}{\ms{center}}\nc{\ce}{\ms{center}}
\nc{\ci}{\circ}
\nc{\code}{\ms{code}}\nc{\cod}{\ms{code}}\nc{\decode}{\ms{decode}}\nc{\encode}{\ms{encode}}
\nc{\de}{:\equiv}
\nc{\dr}{\right}\nc{\ga}{\left}
\nc{\ds}{\displaystyle}
\nc{\ep}{\varepsilon}
\nc{\eq}{\equiv}
\nc{\fib}{\ms{fib}}
\nc{\funext}{\ms{funext}}\nc{\fu}{\ms{funext}}
\nc{\gam}{\gamma}
\nc{\glue}{\ms{glue}}\nc{\gl}{\ms{glue}}
\nc{\happly}{\ms{happly}}\nc{\ha}{\ms{happly}}
\nc{\id}{\ms{id}}
\nc{\ima}{\ms{im}}
\nc{\inc}{\subseteq}
\nc{\ind}{\ms{ind}}
\nc{\inl}{\ms{inl}}
\nc{\inr}{\ms{inr}}
\nc{\isContr}{\ms{isContr}}\nc{\co}{\ms{isContr}}\nc{\iC}{\ms{isContr}}\nc{\ic}{\ms{isContr}}
\nc{\isequiv}{\ms{isequiv}}\nc{\iseq}{\ms{isequiv}}\nc{\ieq}{\ms{isequiv}}
\nc{\ishae}{\ms{ishae}}\nc{\ish}{\ms{ishae}}\nc{\ih}{\ms{ishae}}
\nc{\isProp}{\ms{isProp}}\nc{\prop}{\ms{isProp}}\nc{\iP}{\ms{isProp}}\nc{\ip}{\ms{isProp}}
\nc{\isSet}{\ms{isSet}}\nc{\isS}{\ms{isSet}}\nc{\iss}{\ms{isSet}}\nc{\iS}{\ms{isSet}}\nc{\is}{\ms{isSet}}
\nc{\lam}{\lambda}
\nc{\LEM}{\ms{LEM}}\nc{\lem}{\ms{LEM}}\nc{\LE}{\ms{LEM}}
\nc{\lv}{\lvert}\nc{\rv}{\rvert}\nc{\lV}{\lVert}\nc{\rV}{\rVert}
\nc{\Map}{\ms{Map}}
\nc{\merid}{\ms{merid}}\nc{\meri}{\ms{merid}}\nc{\mer}{\ms{merid}}\nc{\me}{\ms{merid}}
\nc{\na}{\ms{nat}}
\nc{\nn}{\noindent}
\nc{\one}{\mb1}
\nc{\oo}{\operatorname}
\nc{\pd}{\prod}
\nc{\ps}{\mc P}
\nc{\pa}{\ms{pair}^=}
\nc{\ph}{\varphi}
\nc{\ppmap}{\ms{ppmap}}
\nc{\pr}{\ms{pr}}
\nc{\Prop}{\ms{Prop}}
\nc{\qinv}{\ms{qinv}}\nc{\qin}{\ms{qinv}}\nc{\qi}{\ms{qinv}}
\nc{\rec}{\ms{rec}}
\nc{\refl}{\ms{refl}}
\nc{\seg}{\ms{seg}}
\nc{\Set}{\ms{Set}}
\nc{\sm}{\scriptstyle}
\nc{\sms}{\ms s}
\nc{\sq}{\square}
\nc{\suc}{\ms{succ}}\nc{\su}{\ms{succ}}
\nc{\tb}{\textbf}
\nc{\then}{\Rightarrow}
\nc{\tms}{\ms t}
\nc{\tx}{\text}
\nc{\transport}{\ms{transport}}\nc{\tr}{\ms{transport}}
\nc{\two}{\mb2}
\nc{\Type}{\text-\ms{Type}}\nc{\type}{\text-\ms{Type}}\nc{\ty}{\text-\ms{Type}}
\nc{\U}{\mc U}
\nc{\ua}{\ms{ua}}
\nc{\uniq}{\ms{uniq}}
\nc{\univalence}{\ms{univalence}}
\nc{\vide}{\varnothing}
\nc{\ws}{\ms{sup}}
\nc{\zero}{\mb0}

\usepackage{amsmath}
\DeclareMathOperator{\sech}{sech}

\begin{document} 

\begin{titlepage} 
\begin{center} 

{\Huge Solitons in the Korteweg-de Vries Equation}\\[0.5cm] 
\textit{Maximilian Bonehill} and \textit{Guillermo Bueno Herranz}~\\[0.3cm] 
School of Physics and Astronomy~\\[0.3cm]
University of Manchester~\\[0.3cm]

~\\[2cm]

\end{center}
{\Large \textbf{Abstract}}~\\[0.3cm]

We propose a numerical solution to the Korteweg-de Vries (KdV) equation using a Crank-Nicolson scheme, and compare its performance to the Fast Fourier Transform method. The properties and interactions of soliton solutions are further examined. Initial conditions were varied to analyse soliton formation in the resulting system. Performing an L$^2$ error analysis demonstrated consistency between numerical methods of solving the KdV equation and analytical solutions. 

\end{titlepage}
\pagenumbering{gobble} 
\clearpage
\pagenumbering{arabic} 
\setcounter{page}{2} 

\newpage 

\section{Introduction} 

First observed by engineer John Scott Russell as a lone wave propagating along an Edinburgh canal in 1834 \cite{history}, solitons are found in a variety of physical systems. Those which arise from the Korteweg-de Vries (KdV) equation \cite{kdv} are employed to model waves in shallow water, signals in optical fibres and particles in quantum field theory \cite{examples}.  

Solitons are generated by non-linear partial differential equations (PDE) including the non-linear Schrödinger, sine-Gordon and KdV equations \cite{example_eq}. Not all PDE support soliton solutions, but those that do are referred to as being "integrable". 

In order for solitons to be produced by the KdV equation, the effects of the breaking and dispersive terms (shown in Figure 1) in the equation must be balanced. They consequently maintain an unchanging, localised form and a constant speed \cite{properties}. Related to these properties and of physical interest are the infinitely-many conserved quantities in the KdV equation, particularly those which correspond to mass, momentum and energy conservation. Note that the principle of superposition does not govern soliton interactions as their generating equation is non-linear. Instead, complex and varied behaviours are observed depending on the individual properties of the interacting waves. We focus on the properties of solitons generated by the KdV equation.

\section{Theory}

Varied numerical methods are applied to obtain solutions to the KdV equation \footnote{Further mathematical methods may be applied to arrive at related equations (such as the modified KdV, general KdV, Sasa-Satsuma, Hirota-Satsuma and Gardner equations as well as the Lax pair formulation of the KdV \cite{footnote_1}). Methods include using Bäcklund and Miura transformations as well as following Hirota's or the inverse scattering method.}. These are calculated by different computation methods, of which the Crank-Nicolson and Fast Fourier Transform methods are compared for accuracy and computational efficiency in this report. The deviation of these numerical solutions from their analytical counterparts was then determined through the L$^2$ norm. 

\subsection{The KdV Equation}

The KdV equation is conventionally expressed as
\begin{equation}
    u_t + \alpha uu_x + \beta u_{xxx} = 0,
    \label{eq:1}
\end{equation}
where the subscript indicates a partial derivative has been taken with respect to that variable, for example $u_t \equiv \frac{\partial u}{\partial t}$. Note that most constant coefficients have been set to unity; however, $\alpha = 6$ and $\beta = 1$ are often chosen by convention to avoid ungainly numerical factors. 

The first term of Equation \ref{eq:1} dictates the motion of the wave in time, the second is the non-linear, advection term and causes breaking whilst the third term corresponds to wave dispersion (see Figure 1).

\begin{figure}
    \centering
    \includegraphics[width=16cm, height=5cm]{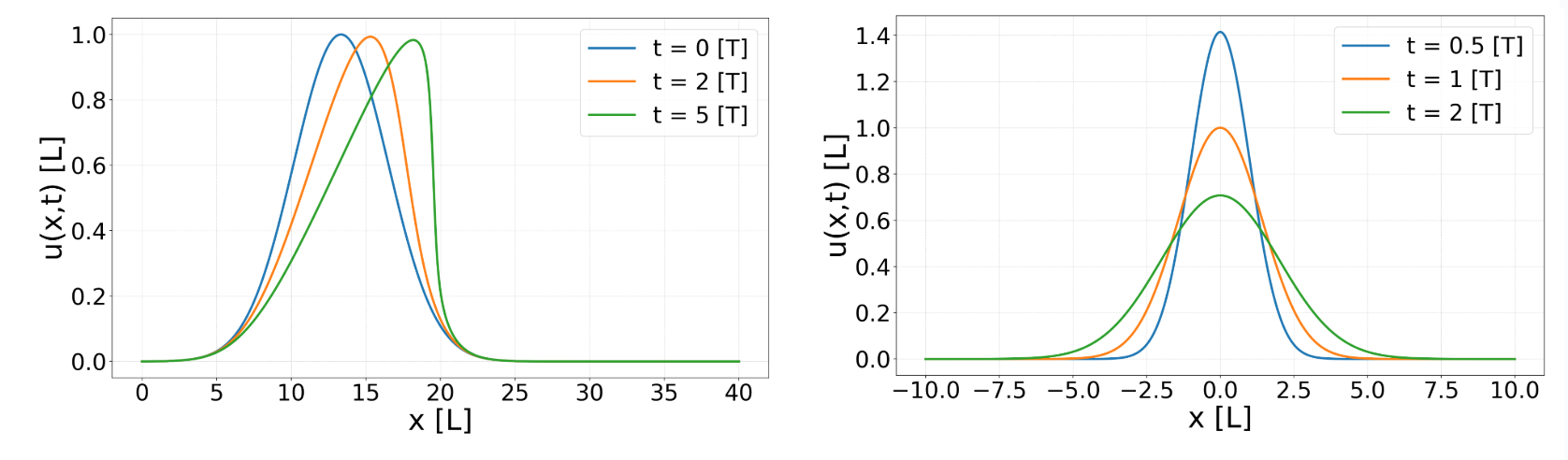}
    \caption{In each illustrated case, the initial condition was taken to be an exact secant-squared soliton solution of the KdV equation with $t = 0$. Excluding the non-linear advection term, dispersion is observed as displayed on the left. In the graph on the right, eliminating the dispersion term illustrates the breaking effect of the advection term. For both graphs, wave height is plotted against position with dimensional units [L] for distance and [T] for time.}
\end{figure}

\subsection{Exact Soliton Solutions of the KdV Equation}

Consider a general solution to the KdV equation of the form $u = f(x-vt)$. Substitute $u$ into Equation~\ref{eq:1} and integrate once to find 
\begin{equation}
    -vf + 3f^2 + f'' = C_1
\end{equation}
where $C_1$ is a constant and $f'$ indicates a total derivative with respect to $x-vt$. Multiplying by $f'$ and integrating once more yields
\begin{equation}
    -\frac{1}{2}vf^2 + f^3 + \frac{1}{2}\left(f'\right)^2 = C_1f + C_2
    \label{eq:3}
\end{equation}
where $C_2$ is another constant. 
If $f$, $f'$ and $f''$ vanish as $x\rightarrow \pm \infty$, $C_1$ and $C_2$ must be zero. Now, rearranging Equation \ref{eq:3} and integrating from $x_0$ at $t=0$ to $x - vt$, we obtain
\begin{equation}
     \int_{}^{} \frac{1}{f\left(v-2f\right)^{\frac{1}{2}}} \,df  = x - x_0 - vt.
\end{equation}
Letting $f = \frac{1}{2}v\sech^2(\theta)$ and evaluating the integral finally gives  
\begin{equation}
    u(x, t) = \frac{1}{2}v\sech^2\left(\frac{\sqrt{v}}{2}(x - x_0 - vt)\right) .
    \label{eq:5}
\end{equation}
This solution to the KdV equation corresponds to a soliton of height $\frac{v}{2}$, width $\frac{1}{\sqrt{v}}$ and speed $v$. 

More solutions to the KdV equation can be found by applying Bäcklund and Miura transforms, combinations of which result in the 2-soliton interaction equation
\begin{equation}
    u(x,t) =\frac{2(c_1 - c_2) \left( c_1 \cosh^2 \left( \frac{\sqrt{c_2} \, \xi_2}{2} \right) + c_2 \sinh^2 \left( \frac{\sqrt{c_1} \, \xi_1}{2} \right) \right)}{\left(\left( \sqrt{c_1} - \sqrt{c_2} \right) \cosh \left( \frac{\sqrt{c_1} \, \xi_1 + \sqrt{c_2} \, \xi_2}{2} \right) + \left( \sqrt{c_1} + \sqrt{c_2} \right) \cosh \left( \frac{\sqrt{c_1} \, \xi_1 - \sqrt{c_2} \, \xi_2}{2} \right) \right)^2}
    \label{eq:6}
\end{equation}

where $v_1$ and $v_2$ are the speeds of the two solitons and $\xi_1=x-x_1-v_1t$ (similarly for $\xi_2$). We categorise two types of interaction by defining a velocity ratio $r = \frac{v_1}{v_2}$. Conventional descriptions dictate if $r > 3$, the solitons appear to "merge then split" while for $r < 3$, they "bounce and exchange" \cite{footnote_1}.

\subsection{Numerical Methods (Crank-Nicolson)}

Numerical methods may be split into two categories, explicit and implicit \footnote{Runge-Kutte, leapfrog and Euler methods can be framed explicitly meanwhile Newton-Raphson, Adams-Bashforth and Crank-Nicolson are examples of implicit methods \cite{explicit/implicit}.}. The former relies on finding the solution at future time-steps using known values of current or previous time-steps, whilst the latter uses both current and future values. The Crank-Nicolson \cite{crank} scheme as presented here is implicit and was employed for its accuracy and property of energy conservation.

\subsubsection{Discretising the KdV Equation}

The method of finite differencing by re-arranging Taylor series was applied throughout to define discrete derivatives. The discretised derivative denoted $u_j^n$ is specified at position $x = j\Delta x$ and time $t = n\Delta t$ where $j\in [0, N-1]$ and $n \in [0,\infty)$. 

The Crank-Nicolson method centres time-related terms at $n+1/2$ and spatial terms at $j+1/2$. Note these points do not explicitly exist in the $\left(j, n\right)$ coordinate grid. 

The first and last terms in the KdV equation are accordingly expressed as 
\begin{equation}
 u_t = \frac{1}{2\Delta t} \left( u_{j}^{n+1} + u_{j+1}^{n+1} - u_{j}^{n} - u_{j+1}^{n} \right)
\label{eq:time_derivative}
\end{equation}
and
\begin{equation}
u_{xxx} = \frac{1}{2\Delta x^{3}} \left( u_{j+2}^{n+1} - 3u_{j+1}^{n+1} + 3u_{j}^{n+1} - u_{j-1}^{n+1} + u_{j+2}^{n} - 3u_{j+1}^{n} + 3u_{j}^{n} - u_{j-1}^{n} \right) .
\label{eq:third_partial}
\end{equation}

We approximate \( \bar{u} \) as the average of \( u_{j} \) and \( u_{j+1} \), though a predictor-corrector scheme is later introduced, as we require \( \bar{u} = u^{n+1/2} \).  
The non-linear term \( u u_x\) is thus 
\begin{equation}
u u_x = \frac{\bar{u}_{j+1} + \bar{u}_{j}}{4\Delta x} \left( u_{j+1}^{n+1} - u_{j}^{n+1} + u_{j+1}^{n} - u_{j}^{n} \right).
\label{eq:nonlinear_term}
\end{equation}
Including the coefficients $\alpha$ and $\beta$, the expressions above are substituted in Equation 1 and terms of $u^{n+1}$ and $u^{n}$ collected. This leads to 
\begin{multline}
 u_{j+2}^{n+1}\left(\frac{\beta}{2\Delta x^{3}}\right) +u_{j+1}^{n+1}\left(\frac{1}{2\Delta t} - \frac{3\beta}{2\Delta x^{3}} + \alpha \frac{\bar{u}_{-} + \bar{u}_{+}}{4\Delta x}\right) +u_{j}^{n+1}\left(\frac{1}{2\Delta t} + \frac{3\beta}{2\Delta x^{3}} - \alpha \frac{\bar{u}_{-} + \bar{u}_{+}}{4\Delta x} \right) +u_{j-1}^{n+1} \left(\frac{-\beta}{2\Delta x^{3}}\right)= \\= u_{j+2}^{n}\left(\frac{-\beta}{2\Delta x^{3}} \right) +u_{j+1}^{n} \left(\frac{1}{2\Delta t} + \frac{3\beta}{2\Delta x^{3}} - \alpha \frac{\bar{u}_{-} + \bar{u}_{+}}{4\Delta x} \right) +u_{j}^{n}\left(\frac{1}{2\Delta t} - \frac{3\beta}{2\Delta x^{3}} + \alpha \frac{\bar{u}_{-} + \bar{u}_{+}}{4\Delta x} \right) +u_{j-1}^{n}\left(\frac{-\beta}{2\Delta x^{3}}\right) .
 \label{eq:10}
\end{multline}
The scheme is more cleanly expressed in matrix form, 
\begin{equation}
\mathbf{A} u^{n+1} = \mathbf{B} u^{n},
\label{eq:11}
\end{equation}
where $\mathbf{A}$ and $\mathbf{B}$ are square matrices of order $N$ while $u^{n+1}$ and $u^{n}$ are treated as column vectors. 

\begin{figure}
    \centering
    \includegraphics[width=0.7\textwidth]{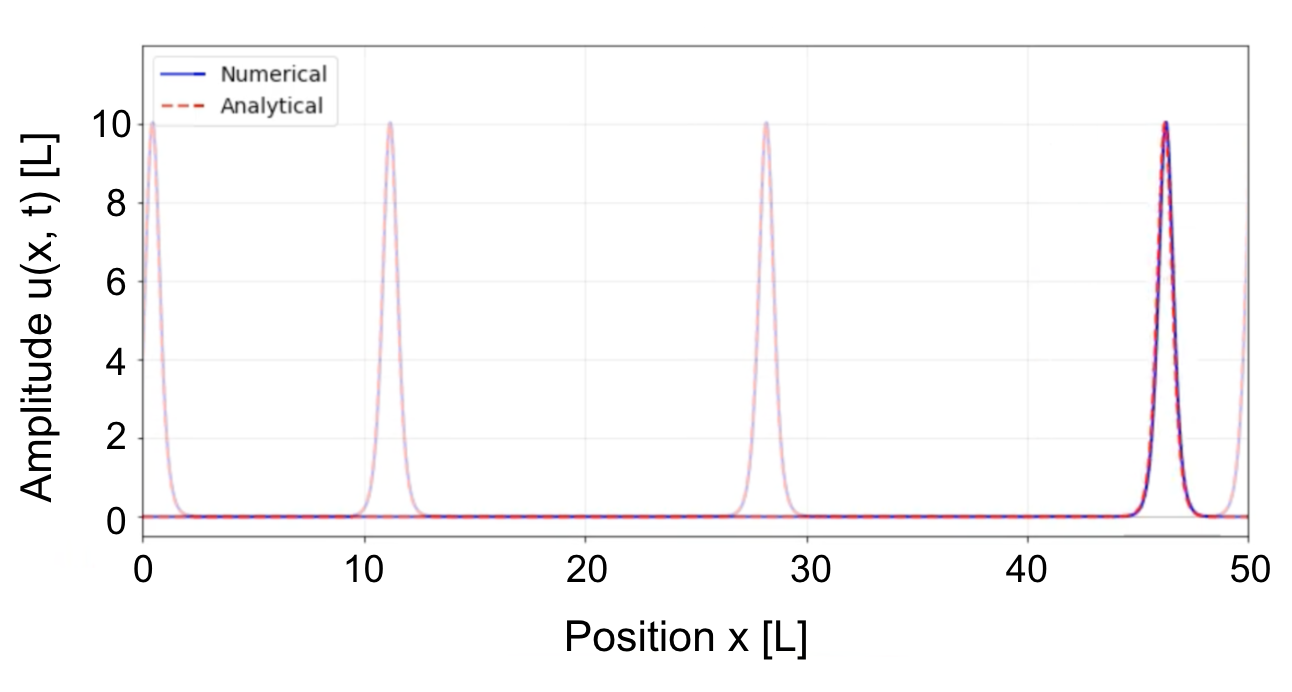}
    \caption{Illustration of periodic boundary conditions applied to the discretised KdV equation for a single soliton evolving to the right from an initial condition of Equation \ref{eq:5} with $t = 0$ on a plot of position against amplitude. Note how the wave wraps around the boundary, appearing to continously re-emerge from the $x = 0$ boundary as it meets that at $x = L$ for the numerical solution as it does for the analytical.}
\end{figure}

As per the analytical solution (Equation \ref{eq:5}), $u \rightarrow 0$ as $x \rightarrow \infty$. However, the modelling domain is finite. As such, periodic boundary conditions were applied (as in Figure 2) for all modelled systems by imposing the condition $u_{j\leq 0}=u_{j+N}$ and $u_{j\geq N}=u_{j-N}$. These fix values of $u$, $u_x$, and $u_{xx}$ at the boundaries. Details on the corresponding matrix coefficients of $\mathbf{A}$ and $\mathbf{B}$ can be found in Appendix A. 

\subsubsection{Predictor-Corrector Technique}
The non-linearity \( \bar{u} \) cannot be directly centred at $n+1/2$ as it is not explicitly a point in the $\left(j, n\right)$ grid, hence a predictor-corrector technique\cite{PC} is implemented.

In each time step, the Crank-Nicolson method is applied twice. In the first step (the predictor step), \( \bar{u} \) is replaced with \( u^n \), the current set of values for \( u \), and the predicted value \( \tilde{u}^{n+1} \) is computed using the Crank-Nicolson method. In the second step (the corrector step), the predicted value is combined with the current value to approximate \( \bar{u}^{n+1/2} \) as
\begin{equation}
\bar{u}^{n+1/2} \approx \frac{u^n + \tilde{u}^{n+1}}{2}.
\end{equation}
The matrices \( \mathbf{A} \) and \( \mathbf{B} \) are rebuilt using this approximation and the Crank-Nicolson method re-applied.
Schematically, each of the steps for the predictor-corrector scheme used follows
\begin{equation}
\begin{split}
&\text{Predictor Step}\;\;\; \bar{u} = u^{n}\; \xrightarrow{\text{Crank-Nicolson}}\; \tilde u^{n+1} = \mathbf{A}(u^{n})^{-1} \left[ \mathbf{B}(u^{n}) u^{n} \right] \\
&\text{Corrector Step}\;\;\; \bar{u}^{n+1/2}= \frac{\tilde u^{n+1} + u^{n}}{2} \; \xrightarrow{\text{Crank-Nicolson}}u^{n+1} = \mathbf{A}(\bar{u}^{n+1/2})^{-1} \left[ \mathbf{B}(\bar{u}^{n+1/2}) u^{n} \right].
\end{split}
\label{eq:predictor_corrector}
\end{equation}

\subsubsection{Stability Analysis of the Crank-Nicolson Scheme}

Performing Von Neumann analysis on the discretised KdV equation reveals its Courant-Friedrich-Lewy (CFL) condition\cite{cfl} which serves as a gauge of the scheme's stability for step sizes $\Delta x$ and $\Delta t$. Formulating the KdV equation with $\frac{\bar{u}_{j+1} + \bar{u}_{j}}{2\Delta x}= \bar{u}$, it follows that 
\begin{equation}
\begin{aligned}
\frac{1}{2\Delta t} &\left( u_{j}^{n+1} + u_{j+1}^{n+1} - u_{j}^{n} - u_{j+1}^{n} \right) + \alpha \frac{\bar{u}}{2\Delta x} \left( \frac{u_{j+1}^{n+1} - u_{j}^{n+1} + u_{j+1}^{n} - u_{j}^{n}}{2} \right)+ \\
&+ \frac{\beta}{2\Delta x^{3}} \left( u_{j+2}^{n+1} - 3u_{j+1}^{n+1} + 3u_{j}^{n+1} - u_{j-1}^{n+1} + u_{j+2}^{n} - 3u_{j+1}^{n} + 3u_{j}^{n} - u_{j-1}^{n} \right) = 0 .
\label{eq:14}
\end{aligned}
\end{equation}
Substituting the Fourier mode $u_j^n = g^n e^{ikj\Delta x}$ into Equation \ref{eq:14} then gives
\begin{align}
\frac{1 + e^{ik\Delta x}}{2\Delta t}& (g^{n+1} - g^n) e^{ikj\Delta x} + \frac{\alpha \bar{u} (e^{ik\Delta x} - 1)}{4\Delta x} (g^{n+1} + g^n) e^{ikj\Delta x} +\\
&+\frac{\beta (e^{2ik\Delta x} - 3e^{ik\Delta x} + 3 - e^{-ik\Delta x})}{2\Delta x^3} (g^{n+1} + g^n) e^{ikj\Delta x} = 0 .
\end{align}
Now collecting factors of $g^{n+1}$ and $g^{n}$,
\begin{equation}
\left[ \frac{1 + e^{ik\Delta x}}{2\Delta t} + \frac{\alpha \bar{u} (e^{ik\Delta x} - 1)}{4\Delta x} + \frac{\beta A(k)}{2\Delta x^3} \right] g^{n+1} = \left[ \frac{1 + e^{ik\Delta x}}{2\Delta t} - \frac{\alpha \bar{u} (e^{ik\Delta x} - 1)}{4\Delta x} - \frac{\beta A(k)}{2\Delta x^3} \right] g^n
\end{equation}

where $A(k) = e^{2ik\Delta x} - 3e^{ik\Delta x} + 3 - e^{-ik\Delta x}$ for convenience. Hence the amplification factor $g$, that is, the ratio of $g^{n+1}$ to $g^{n}$, must be
\begin{equation}
g =\frac{
    2(1 + e^{ik\Delta x})\Delta x^3 - (e^{ik\Delta x} - 1)\alpha \bar{u} \Delta x^2 \Delta t - 2A(k)\beta \Delta t
}{
    2(1 + e^{ik\Delta x})\Delta x^3 + (e^{ik\Delta x} - 1)\alpha \bar{u} \Delta x^2 \Delta t + 2A(k)\beta \Delta t
} .
\end{equation}
For the scheme to be stable, it is required that $|g| \leq 1$. Simplifying the expression, we obtain
\begin{equation}
    |g|=1 
\end{equation}
which demonstrates that the scheme is stable, meaning the solution will no blow up, for any combination of step sizes $\Delta x$ and $\Delta t$. 

\subsubsection{Fast Fourier Transform (FFT) Applied to the KdV Equation}

Fast Fourier transforms (FFT)\cite{fft} are used to rapidly compute the discrete Fourier transform of a sequence of values. Given points $x_n = x_0 + j \Delta x$ with $j \in [0, N-1]$, the FFT brings them into the frequency space with $\Delta k = 2 \pi / (N \Delta x)$ as  
\begin{equation}
k_j =
    \begin{cases}
        j \Delta k, & \text{if } 0 \leq j \leq \frac{N}{2}\\
        \left(j-N\right)\Delta x, & \text{if } \left(\frac{N}{2} + 1\right) \leq j \leq \left(N - 1\right)
    \end{cases}.
\end{equation}
Fourier transforming the KdV, with transformed terms indicated by hat notation, results in 
\begin{equation}
   u_t+\frac{\alpha}{2}\left( u\right) ^{2}_x+u_{xxx}=0\; \xrightarrow{\text{Fourier Transform}}\; \hat{u}_t = - i\frac{\alpha}{2}k\widehat{\left(u^2\right)} + i\beta k^3\hat{u} .
\end{equation}
Split-stepping, the left side in the Fourier transformed expression is equated to each of the right side terms individually. This enables the first to be solved for one time step and to then be substituted into the second,
\begin{equation}
\displaystyle{ \begin{matrix}
\hat{u}_{a}\left( k,t+\Delta t\right)  &=&\hat{u}\left( k,t\right)
\mathrm{e}^{\mathrm{i}k^{3}\Delta t} \\
\hat{u}_b\left( k,t+\Delta t\right)  &=&\hat{u}_{a}\left( k,t+\Delta t\right)
- 3ik\Delta t\left( \mathcal{F}\left( \left( \mathcal{F}^{-1}\left[ \hat{u}_{1}\left( k,t+\Delta t\right)\right]
\right) ^{2}\right) \right) 
\end{matrix} },
\end{equation}
thus iteratively solving the KdV equation.

Note that implementing periodic boundary conditions in the Crank-Nicolson method enables comparison with the FFT method which is inherently periodic.

\subsubsection{Identifying Soliton Solutions}

In order to identify soliton solutions, two points are selected at half-amplitude either side of the wave crest. The distance between these points along with the velocity and maximum amplitude of the wave are calculated for all points in time. If these measures tend towards constant values, then we deduce that the observed wave has the basic properties of a soliton. Plots of these measures are included in Figure 3. 
\begin{figure}
    \centering
     \makebox[\textwidth]{\includegraphics[width=1.15\linewidth]{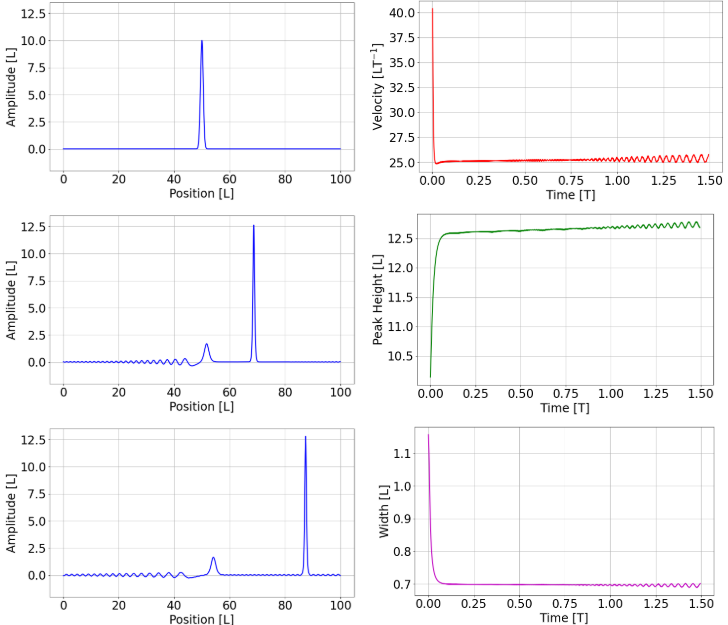}}
    \caption{Down the left side, a Gaussian initial condition is seen evolving a primary wave and secondary wavelets travelling in the opposite direction for times t = 0.000, 0.740 and 1.480 [T] on a plot of amplitude against position. Down the right side, peak height, width and velocity tend towards constant average values following the decay of transient behaviour. The oscillating pattern results from interactions between the wavelets and the soliton as they pass the boundary with periodic conditions and intersect. Note $\Delta t = 10^{-3}$ [T] and $\Delta x = 0.05$ [L].} 
\end{figure}
\subsubsection{Error Analysis}

Methods of finite differencing involve truncated power series expansions, in this case involving (non-zero) time and space intervals ($\Delta t$ and $\Delta x$), so they necessarily contain an associated error. A measure of the deviation of the numerical solution \( u_{\text{num}} \) from the corresponding analytical solution \( u_{\text{exact}} \), that is an evaluation of the quality of the numerical approximation, for each position $\left(j, n\right)$ is provided by the discrete analogue of the $L^2$ norm. The expression for the local $L_x^2$ is given by 
\begin{equation}
\|u^n\|_{L^2_x}= \sqrt{\Delta x  \sum_{j=1}^{N_x} |u_{\text{exact}}(x_j, t_n) - u_{\text{num}}(x_j, t_n)|^2}.
\end{equation}
Whilst the error associated with the scheme is inherently local, the accumulation of these errors over the total time across which the simulation runs (and so across all discrete steps) results in a global error. The revised $L^2$ follows from averaging the $L_x^2$ over time (analogous to the Bochner space norm), 
\begin{equation}
\|u\|_{L^2_{x,t}} = \sqrt{\frac{\Delta x \Delta t}{T} \sum_{n=1}^{N_t} \sum_{j=1}^{N_x} |u_{\text{exact}}(x_j, t_n) - u_{\text{num}}(x_j, t_n)|^2}.
\end{equation}
It should be noted that $\Delta x=l/N_x$ where $l$ represents the length of the domain spanned by the solution and $N_x$ is the number of grid divisions within the domain equivalent to the number of possible $j$. Similarly, $\Delta t=T/N_t$ where $T$ is the total time over which the numerical solution evolves and $N_t$ corresponds to the number of points in time as does the number of values of $n$. As only the global error will be further discussed, we re-notate $L^2 =\|u\|_{L^2_{x,t}}$.

\subsubsection{Error Convergence}

Noting the scheme is of order $O(\Delta x^2) + O(\Delta t^2)$, a convergence analysis of the numerical scheme was performed by fixing $\Delta x$ at $5.00\times 10^{-3}$ [L] while varying $\Delta t$ from $10^{-4}$ to $10^{-2}$. A similar test was performed holding $\Delta t$ constant whilst varying $\Delta x$. Both norms were found to converge at the expected rates expressed according to 
\begin{equation}
    \text{Rate} = \log\left(\frac{\text{Error}_i}{\text{Error}_{i+1}}\right) \left(\log\left(\frac{N_{i+1}}{N_i}\right)\right)^{-1}
\end{equation}
where $i$ and $i+1$ denote subsequent runs of the convergence analysis. Calculated global $L^2$ errors are consequently interpreted as measures of cumulative deviation only for a specific $\Delta t$ and $\Delta x$ pair. As global L$^2$ error analysis agreed with the order of the scheme, consistency between the analytical and numerical solutions was achieved.

\newpage

\section{Results}
Note that times, distances and speeds were proportioned to create suitable figures (just as $\alpha$ and $\beta$ are selected in the KdV equation according to the desired scaling), but that dimensional units of [T], [L] and [LT$^{-1}$] are employed as no calculation for data with specific units is being performed. 

\subsection{1-Soliton Systems}

Setting a Gaussian initial condition, 
\begin{equation}
u(x,t) = 10 \exp\left(-\frac{1}{2}{\left(x - \frac{l}{2}\right)^2}\right),
\end{equation}
resulted in the production of a primary wave as well as an oscillatory pattern of secondary waves (displayed in Figure 3). The latter were emitted from the tail of the primary wave and progressed in the opposite direction, possessing much smaller amplitudes which varied above and below the horizontal axis. These secondary waves are physically interpreted as a conservation effect, perhaps containing energy that is shed from the primary wave which narrows in width during its transient phase before becoming stable, at which point no more secondary waves are produced. 

The distance between points on the primary wave envelope and the wave's height and velocity tended towards a constant average value following a decay of transient behaviour visible in Figure 3 up to approximately 0.1 [T]. The wave thus evolved into behaving as a soliton given a Gaussian initial condition. 

Due to periodic boundary conditions, the secondary waves of increasing size wrapped around the boundary then interfered with the primary wave. The effect is a growing oscillation in the soliton-determining measures (seen in Figure 3). The primary wave's return to its initial state supports its soliton nature. 

\subsection{2-Soliton Systems}

\begin{figure}
    \centering
    \includegraphics[width=0.7\textwidth]{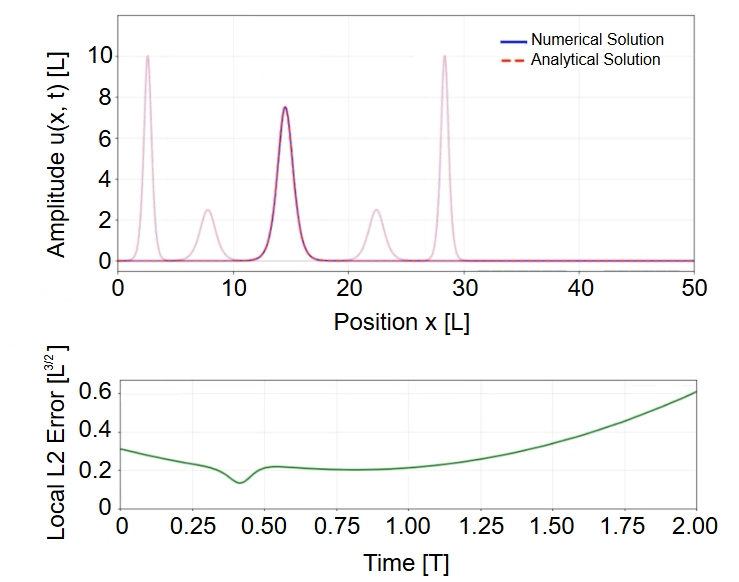}
    \caption{Shown above in a plot of position versus amplitude is a merge-split interaction, corresponding to $r > 3$, between two solitons evolving to the right from a secant-squared initial condition (that is, Equation \ref{eq:5} with $t=0$). Step sizes of $\Delta x = 0.025$ and $\Delta t = 10^{-3}$ were used with a total error of 0.310 in the numerical solution compared to the analytical solution. A graph of local L$^2$ error recorded as a function of time is displayed. Note the dip which occurs during the interaction. The analytical solution was modelled by Equation \ref{eq:6}.}
\end{figure}

Two cases of 2-soliton interaction were observed corresponding to merge-split and bounce-exchange interactions. Measures of height, speed and width were performed as for the Gaussian initial condition. It was found that each measure remained constant in both cases before and after the interaction, confirming the waves behaved as expected of solitons. 

The merge-split interaction (see Figure 4) required a velocity ratio $r > 3$ as expected. During this interaction, the waves slid into one another, then uncombined and returned to their original forms. A global error of 0.310 [L$^\frac{3}{2}$] was calculated. As the profile merges, the envelope's length reaches a minimum, reducing the space over which the L$^2$ error accumulates and producing a dip in the L$^2$ graph.

\begin{figure}
    \centering
    \includegraphics[width=0.7\textwidth]{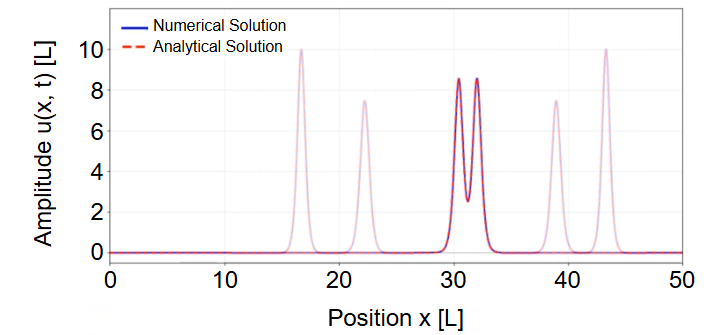}
    \caption{Above is a bounce-exchange 2-soliton interaction, indicating $r < 3$, on a plot of amplitude against position. Each soliton was produced with a secant-squared initial condition (from Equation \ref{eq:5} with $t=0$) with a velocity in the positive x-direction, then stepped by $\Delta x = 0.025$ and $\Delta t = 10^{-3}$ to arrive at a global error of 0.336 [L$^\frac{3}{2}$]. The analytical solution was plotted using Equation \ref{eq:6}.}
\end{figure}

With $r < 3$, a bounce-exchange interaction was observed (shown in Figure 5) whereby the two waves sidled up to one another and appeared to exchange energy before decoupling and progressing as before. A smaller dip in the L$^2$ trend than that in the merge-split interaction was observed. The larger L$^2$ and global error, in this case 0.336 [L$^\frac{3}{2}$], may be attributed to the complex profile observed during the interaction. 

In the case of $r = 3$, a balanced combination of each interaction behaviour was observed. The individual solitons evolved into one profile just as they assumed the same amplitude, then smoothly demerged. 

\newpage

\subsection{Comparing Crank-Nicolson and FFT Results}

\begin{figure}
    \centering
    \includegraphics[width=0.7\textwidth]{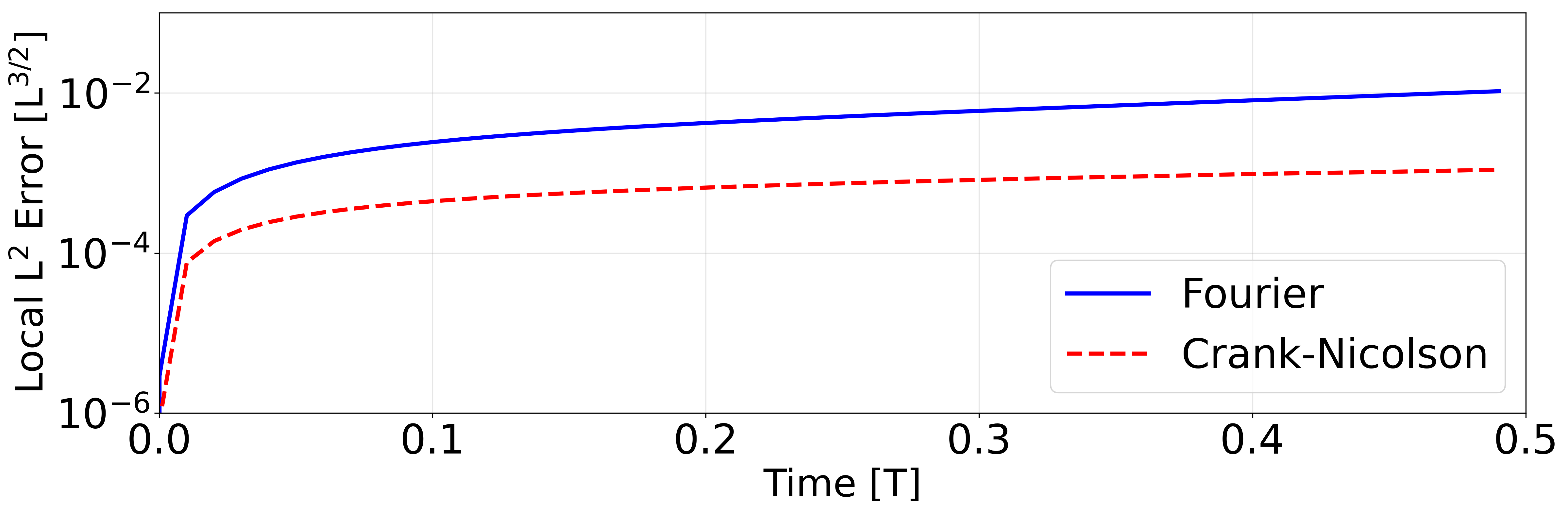}
    \caption{Graph of local L$^2$ error accumulated over time by the Crank-Nicolson and FFT methods when applied to solve the 1-soliton system with $\Delta t = 10^{-4}$ and $\Delta x \approx 9.8 \times 10^{-2}$. Note step sizes were chosen to maximise the FFT efficiency. The error in both methods gathers on a logarithmic scale.}
\end{figure}

Numerical solutions acquired by both the Crank-Nicolson and FFT methods were applied to model a 1-soliton system. When evolving the system, the error developed on a logarithmic scale in both cases. While the Crank-Nicolson method accumulated a global error of $7.43 \times 10^{-5}$, the FFT method grew a much larger error of $5.85 \times 10^{-3}$ (as seen in Figure 6). The first method, however, took approximately 100 times longer to run than the FFT, so although more accurate, the Crank Nicolson method is more computationally expensive.

\section{Conclusions and Extensions}

Periodic boundaries were applied to evolve systems of 1- and 2-solitons generated by the KdV equation using Crank-Nicolson and FFT numerical methods. Applying a Gaussian initial condition resulted in the observation of a wave with soliton properties following a decay in transient behaviour. Setting two solitons in motion in the system resulted in two species of interaction; that which occurred depended on the ratio of the speeds of the individual solitons. 

Comparing computational methods, it was found that the Crank-Nicolson method produced more accurate results than the FFT method, though the latter generated results far more rapidly.

The project lends itself to myriad extensions, the most accessible of which involve varying boundary conditions (for example to model absorption, reflection and transmission, perhaps as it occurs at an interface of media of different refractive indices) and initial conditions (to further investigate the way in which a wave may evolve towards presenting as a soliton solution). Related equations, such as a generalised KdV (gKdV) of the form $ u_t + \alpha u^p u_x + \beta u_{xxx} = 0$ where, conventionally, $p \in \Z^+$ may also be explored. Setting $p = 1/2$, for example, the gKdV resembles the Schamel equation which models plasma behaviour with its soliton solutions. Additionally, N-soliton systems where $N > 2$ may be modelled and the response in accuracy of the numerical method (Crank-Nicolson, FFT or another) may be investigated. Finally, a variety of numerical methods may be examined for considerations including computing time, graphic quality and solution accuracy. 

\bibliography{bibliography} 

\newpage
\appendix

\section{Matrix coefficients}

The non-zero elements of \( \mathbf{A} \) and \( \mathbf{B} \) are
\begin{equation}
\begin{aligned}
a_{j,j-1} , \;\;\;\;a_{j,j}, \;\;\;\;a_{j,j+1}, \;\;\;\;a_{j,j+2}, \;\;\;\; b_{j,j-1}, \;\;\;\;b_{j,j}, \;\;\;\;b_{j,j+1}, \;\;\;\;b_{j,j+2} 
\end{aligned}
\label{eq:matrix_elements}
\end{equation}
using the coefficients of Equation \ref{eq:10}
\begin{equation}
\begin{aligned}
a_{j,j-1} &= -\frac{\beta}{2\Delta x^{3}} \;\;\;\; a_{j,j} = \frac{1}{2\Delta t} + \frac{3\beta}{2\Delta x^{3}} - \alpha \frac{\bar{u}_{-} + \bar{u}_{+}}{4\Delta x} \;\;\;\; a_{j,j+1} = \frac{1}{2\Delta t} - \frac{3\beta}{2\Delta x^{3}} + \alpha \frac{\bar{u}_{-} + \bar{u}_{+}}{4\Delta x} \;\;\;\; a_{j,j+2} = \frac{\beta}{2\Delta x^{3}} \\
b_{j,j-1} &= \frac{\beta}{2\Delta x^{3}} \;\;\;\;b_{j,j} = \frac{1}{2\Delta t} - \frac{3\beta}{2\Delta x^{3}} + \alpha \frac{\bar{u}_{-} + \bar{u}_{+}}{4\Delta x} \;\;\;\;b_{j,j+1} = \frac{1}{2\Delta t} + \frac{3\beta}{2\Delta x^{3}} - \alpha \frac{\bar{u}_{-} + \bar{u}_{+}}{4\Delta x} \;\;\;\;b_{j,j+2}= -\frac{\beta}{2\Delta x^{3}}
\end{aligned}
\label{eq:matrix_coefficients}
\end{equation}

where \( \bar{u}_{-} = u_{j} \) and \( \bar{u}_{+} = u_{j+1} \).

In the case of periodic boundaries where we impose conditions $u_{j\leq 0}=u_{j+N}$ and $u_{j\geq N}=u_{j-N}$, The matrices \( \mathbf{A} \) and \( \mathbf{B} \) (as in Equation \ref{eq:11}) are, using the notation above, structured like matrix \( \mathbf{M} \) 
\begin{equation}
\mathbf{M} =
\begin{bmatrix}
m_{0,0} & m_{0,1} & m_{0,2} & 0 & \dots & 0 & m_{0,-1} \\
m_{1,0} & m_{1,1} & m_{1,2} & m_{1,3} & \dots & 0 & 0 \\
0 & m_{2,1} & m_{2,2} & m_{2,3} & \dots & 0 & 0 \\
\vdots & \vdots & \vdots & \vdots & \ddots & \vdots & \vdots \\
m_{N-2,N} & 0 & \dots & 0 & m_{N-2,N-3} & m_{N-2,N-2} & m_{N-2,N-1} \\
m_{N-1,N} & m_{N-1,N+1} & 0 & \dots & 0 & m_{N-1,N-2} & m_{N-1,N-1}
\end{bmatrix}.
\label{eq:matrix_ref}
\end{equation}
The non-zero elements of matrix \( \mathbf{M} \) appear to wrap around the matrix just as the soliton wraps around the boundary.

\end{document}